\documentclass[12pt]{article}
\usepackage{amsmath}
\usepackage{graphicx}%
\usepackage{amsfonts}%
\usepackage{amssymb}

\newcommand{\bea}{\begin{eqnarray}}
\newcommand{\eea}{\end{eqnarray}}
\newcommand{\be}{\begin{equation}}
\newcommand{\ee}{\end{equation}}
\newcommand{\vs}[1]{\vspace{#1 mm}}

\newcommand{\dsl}{\pa \kern-0.5em /}

\newcommand{\pa}{\partial}

\newcommand{\nn}{\nonumber\\}

\newcommand{\eqn}[1]{(\ref{#1})}

\begin{document}
\topmargin 0pt
\oddsidemargin 0mm

\begin{flushright}

USTC-ICTS-09-13\\

\end{flushright}

\vspace{2mm}

\begin{center}

{\Large \bf Holographic Superconductor  for a  Lifshitz fixed point}
\vs{6}

{\large  Sang-Jin Sin\footnote{E-mail:sangjin.sin@gmail.com}, Shan-Shan
Xu\footnote{E-mail: xuss@mail.ustc.edu.cn}, Yang Zhou\footnote{E-mail:yzhou@itp.ac.cn} }

\vspace{4mm}

{\em

 1)Department of Physics, Hanyang University, Seoul 133-791, Korea

 2)Interdisciplinary Center for Theoretical Study\\

 University of Science and Technology of China, Hefei, Anhui
 230026, China\\

 3)Institute of Theoretical Physics, Chinese Academic of Science, Beijing 100190, PRC\\

Kavli Institute for Theoretical Physics China at the Chinese Academy of Sciences(KITPC-CAS)\\}

\vs{4}

\end{center}


\begin{abstract}
We consider the gravity dual of strongly coupled system at a
Lifshitz-fixed point and finite temperature, which was constructed
in a recent work arXiv:0909.0263. We construct an Abelian Higgs
model in that background and calculate condensation and conductivity
using holographic techniques. We find that condensation happens and
DC conductivity blows up when temperature turns below a critical
value.

\end{abstract}
\newpage

\tableofcontents

\bigskip

\section{Introduction}
AdS/CFT correspondence \cite{Maldacena:1997re} is one of the most
interesting results  in the sense that it opened a window of
connecting the
string theory to the QCD and condensed matter systems. %
The connection between gauge theory and strings has a long history
since the appearance of string models of hadrons in 1960's:  For
example, it has been observed that the elementary excitations of a
lattice gauge theory in the strong coupling limit can be represented
by strings formed by color-electric fluxes. It is also suggested
that in a certain limit all the degrees of freedom in the gauge
theory should be represented by the flux lines (strings) instead of
fields. See \cite{Wilson:1974},\cite{Polyakov:2001af} and
references therein. Therefore it is natural to expect an exact
duality between gauge fields and strings although its precise
formulation was obtained
 only recently.
The semi-classic version of this duality can be stated as gauge/gravity duality  and such duality has become a powerful tool to understand the strongly coupled   QCD and the properties of quark gluon plasma in heavy ion collisions
at RHIC \cite{son,SZ,Mateos:2007ay} as well as the low energy hadron physics.

More recently, it has been attempted to use this correspondence  to
describe  certain condensed matter systems  such as the Quantum Hall
effect~\cite{Hartnoll:2007ai} , Nernst effect~\cite{Hartnoll:2007ih,
Hartnoll:2007ip, Hartnoll:2008hs},
superconductor~\cite{Hartnoll:2008vx,Ammon:2009fe,Horowitz:2008bn}
and fractional quantum hall effect (FQHE)~\cite{Fujita:2009kw}.
These phenomena were suggested to have dual gravitational
descriptions. As pointed out in~\cite{Herzog:2008wg}, there is a
large class of interesting strongly correlated electronic and atomic
systems that can be created and studied in experiments and there are
non-relativistic systems which have Schr$\ddot{\textrm{o}}$dinger
symmetry~\cite{Herzog:2008wg}. However,  the dynamics of such
systems near a critical point is described by a relativistic
conformal field theory or sometimes more subtle scaling theory
having Lifshitz symmetry~\cite{Kachru:2008yh}.

 To describe the finite temperature version of such scaling system,
 four dimensional  black hole solutions with asymptotically Lifshitz spacetimes were
investigated~\cite{Mann:2009yx,Bertoldi:2009vn,Bertoldi:2009dt,Danielsson:2009gi,Brynjolfsson:2009ct}.
The Lifshitz black holes in arbitrary dimensions were also found in
a different class of action~\cite{Taylor:2008tg}. Recently  an
analytical solution
 in yet another  action   was proposed for
z=2~\cite{Balasubramanian:2009rx} in four dimension. Additionally,
Lifshitz black holes in three-dimensional massive gravity and
four-dimensional $R^2$ gravity were also
discussed~\cite{AyonBeato:2009nh,Cai:2009ac}. Embedding those black
holes with the action in~\cite{Kachru:2008yh} into string theory was
addressed in~\cite{Li:2009pf}.


In~\cite{Hartnoll:2008vx}, a model of a strongly coupled system
which shows superconductivity was constructed based on holography,
which is an Abelian-Higgs model in a warped space time. While the
electrons in real materials are non-relativistic, the model
in~\cite{Hartnoll:2008vx} is relativistic. Therefore it is natural
to ask whether one can develop a similar model with non-relativistic
kinematics~\cite{Pu:2009wn}, especially at Lifshitz-like fixed
point. One purpose of this paper is to answer this question. We find
that there is  a critical temperature, like the relativistic case,
below which a charged scalar field condensate and the (DC)
conductivity blows up.
We also calculated the frequency dependent conductivity.

This paper is organized as follows. In section \ref{equilibrium}, we
check thermodynamics of the Lifshitz black hole and chemical
potential background. In the following section, we study the
superconductive phases in the Lifshitz background. We obtain similar
results as in the usual AdS black hole background.

\section{Gravity dual  of the Lifshitz fixed point}\label{equilibrium}
We begin with the equilibrium properties of the strongly coupled
thermal field at Lifshitz-like fixed point, by analyzing the
Lifshitz black hole solutions.
\subsection{Lifshitz black hole solutions}
The Lifshitz scaling is defined by
\begin{equation} \label{5eq1}
t\rightarrow\lambda^z t,\,\,\, x\rightarrow\lambda x,
\end{equation}
where $z$ is called dynamical exponent. The metric with this
symmetry was first found in \cite{Kachru:2008yh}:
\begin{equation} \label{5eq2}
ds^2=L^2\left(-\frac{dt^2}{r^{2z}}+\frac{dx^2+dy^2}{r^2}+\frac{dr^2}{r^2}\right),
\end{equation}
where $0<r<\infty$ and $L$ sets the scale for the radius of
curvature. For $z=1$, this geometry is anti-de sitter
spacetime.  
For $z>1$, it is a candidate for the dual gravity of a field theory
with Lifshitz scaling. 

The tidal
forces diverge on the ``horizon" at $r\rightarrow\infty$ unless
$z=1$ and this implies that the metric \eqn{5eq2} has no global
extension~\cite{Hartnoll:2009sz}. To describe the physics of the
dual field theory at finite temperature, black hole solutions with
asymptotical Lifshitz metric \eqn{5eq2} were studied
in~\cite{Mann:2009yx,Bertoldi:2009vn,Bertoldi:2009dt,Danielsson:2009gi,Brynjolfsson:2009ct}.

Recently, a 4D black hole solution which asymptotes to the
Lifshitz spacetime \eqn{5eq2} was
constructed\cite{Balasubramanian:2009rx}. The action is
\begin{equation} \label{5eq4}
S=\frac{1}{2}\int d^4x\sqrt{-g}\left(R-2\Lambda\right)-\int
d^4x\sqrt{-g}\left(\frac{e^{-2\phi}}{4}F_{\mu\nu}F^{\mu\nu}+\frac{m^2}{2}A_{\mu}A^{\mu}+(e^{-2\phi}-1)\right),
\end{equation}
where $\Lambda=-\frac{z^2+z+4}{2}$, $m^2=2z$ and $F=dA$. The
gravitational constant and curvature radius are set by $8\pi G_4=1$
and $L=1$ respectively. With this convention the equations of motion
  are
\begin{eqnarray} \label{5eq0}
F^2&=&-4,\,\,\,\,\,\,\frac{1}{\sqrt{-g}}\partial_{\mu}\left(\sqrt{-g}e^{-2\phi}F^{\mu\nu}\right)=m^2A^{\nu},\nn
R_{\mu\nu}&=&e^{-2\phi}F_{\mu\lambda}F_{\nu}^{\lambda}+m^2A_{\mu}A_{\nu}+\Lambda
g_{\mu\nu}+\left(2e^{-2\phi}-1\right)g_{\mu\nu},
\end{eqnarray}
and the  black hole solution of this system is \footnote{There is a
factor $1/\sqrt{2}$ missing in the expression (2.5) of the massive
vector field in \cite{Balasubramanian:2009rx}.}
\begin{eqnarray} \label{5eq5}
ds^2=-f(r)\frac{dt^2}{r^{2z}}+\frac{dx^2+dy^2}{r^2}+\frac{dr^2}{r^2f(r)},\nn
f(r)=1-\frac{r^2}{r_H^2},\,\,\,\,\,e^{-2\phi}=1+\frac{r^2}{r_H^2},\,\,\,\,\,A=\frac{f(r)}{\sqrt{2}r^2}dt.
\end{eqnarray}
In the rest of this paper, we will use this solution to discuss the
transport and superconductivity.

\subsection{Thermodynamics}
We first review the thermodynamics of this black hole proposed in
\cite{Balasubramanian:2009rx}. Our calculational procedure  follows
\cite{Hartnoll:2009sz}. According to the AdS/CFT dictionary,
the partition function of the bulk theory is
identified to that of the dual field theory.
 The path integral over metrics is dominated by the saddle point $g_*$,
and the partition function is
\begin{equation} \label{5eq8}
Z=e^{-S_E[g_*]},
\end{equation}
where $S_E[g_*]$ is the Euclidean action evaluated on the saddle.
This action must contain extrinsic boundary terms and intrinsic
boundary terms in order to render the finiteness of the on-shell
action.
This was already given in
\cite{Balasubramanian:2009rx}:
\begin{eqnarray} \label{5eq9}
S_E&=&-\frac{1}{2}\int d^4x\sqrt{g}\left(R-2\Lambda\right)+\int
d^4x\sqrt{g}\left(\frac{e^{-2\phi}}{4}F^2+\frac{m^2}{2}A^2+(e^{-2\phi}-1)\right)\nn
& &+\int_{r\rightarrow 0}d^3x
\sqrt{\gamma}K-\frac{1}{2}\int_{r\rightarrow 0}d^3x
\sqrt{\gamma}\left(-\frac{27}{8}+\frac{7}{2}\phi+\frac{7}{2}\phi^2\right)\nn
& &-\frac{1}{2}\int_{r\rightarrow 0}d^3x
\sqrt{\gamma}\left(\left(\frac{17}{2}+7\phi\right)A^2+\frac{13}{2}A^4\right),
\end{eqnarray}
where $\gamma$ is the induced metric on the boundary $r\rightarrow
0$ and $K$ is the trace of the extrinsic curvature. The Dirichlet
boundary condition is imposed on the massive vector
field.\footnote{That is $c_N$=0 in the expression (3.3) in
\cite{Balasubramanian:2009rx}. We rewrote that expression into
Euclidean space and substitute the specific values of $c_0$-$c_5$.
There is a minus sign difference of $c_1$-$c_5$ here from those
given in Appendix A of \cite{Balasubramanian:2009rx}.}

A saddle is obtained by Wick rotating eq. \eqn{5eq5}:
\begin{eqnarray} \label{5eq6}
ds^2_*=f(r)\frac{d\tau^2}{r^{2z}}+\frac{dx^2+dy^2}{r^2}+\frac{dr^2}{r^2f(r)},\,\,\,\,\,A=-i\frac{f(r)}{\sqrt{2}r^2}d\tau .
\end{eqnarray}
The temperature of the system is
\begin{eqnarray} \label{5eq7}
T=\frac{1}{\beta}=\frac{1}{2\pi r_H^z},
\end{eqnarray}
determined  by the absence of the conical singularity at
$r=r_H$.

We can now evaluate the action \eqn{5eq9}:
\begin{equation} \label{5eq10}
S_E[g_*]=-\beta\frac{L_xL_y}{2r_H^4}=-2\pi^2L_xL_yT,
\end{equation}
and the free energy
\begin{equation} \label{5eq11}
\mathcal{F}=-T\log
Z=TS_E[g_*]=-\frac{L_xL_y}{2r_H^4}=-2\pi^2L_xL_yT^2,
\end{equation}
as given in~\cite{Balasubramanian:2009rx}.

As a check, the entropy
\begin{equation} \label{5eq12}
\mathcal{S}=-\frac{\partial\mathcal{F}}{\partial T}=4\pi^2L_xL_yT,
\end{equation}
coincides with the Bekenstein-Hawking entropy $\mathcal{S}=2\pi A$
with the area of the event horizon $A=L_xL_y/r_H^2$ and the unit
convention $8\pi G=1$.

The boundary stress tensor resulting from \eqn{5eq9}
is\footnote{There are also some minus sign differences between the
expression we give and the one in (3.4) of
\cite{Balasubramanian:2009rx}.}
\begin{eqnarray} \label{5eq13}
T_{\mu\nu}&\equiv&-\frac{2}{\sqrt{-\gamma}}\frac{\delta
S}{\delta\gamma^{\mu\nu}}=K_{\mu\nu}-\left(\frac{17}{2}+7\phi+13A^2\right)A_{\mu}A_{\nu}-K\gamma_{\mu\nu}\nn
&
&+\frac{1}{2}\left(-\frac{27}{8}+\frac{7}{2}\phi+\frac{7}{2}\phi^2\right)
\gamma_{\mu\nu}
+\frac{1}{2}\left(\left(\frac{17}{2}+7\phi\right)A^2+\frac{13}{2}A^4\right)\gamma_{\mu\nu},
\end{eqnarray}
then the internal energy and pressure of boundary theory following
\cite{Balasubramanian:2009rx} are respectively
\begin{eqnarray} \label{5eq14}
\mathcal{E}&=&-L_xL_y\sqrt{-\gamma}T_t^t=\frac{L_xL_y}{2r_H^4},\nn
\mathcal{P}&=&\frac{1}{2}L_xL_y\sqrt{-\gamma}T_i^i=L_xL_y\sqrt{-\gamma}T_x^x=\frac{L_xL_y}{2r_H^4}.
\end{eqnarray}
Thus we have
\begin{equation} \label{5eq15}
\mathcal{E}=\mathcal{P}=-\mathcal{F}=\frac{1}{2}T\mathcal{S}.
\end{equation}
The first law of thermodynamics
$\mathcal{E}+\mathcal{P}=T\mathcal{S}$ is satisfied as given in
\cite{Balasubramanian:2009rx}.

\subsection{Finite chemical potential}
Now we consider a probe gauge field fluctuation $A_\mu$ in the
Lifshitz black hole background. That means we need to add another
Maxwell term to the original action \eqn{5eq4}~\footnote{In this
subsection and the rest of this paper, $A$ and $F$ should be distinguished with those in the
original action \eqn{5eq4} where the vector filed is not a gauge field. From now on, by $A_\mu$ and $F$ we mean the Maxwell fluctuation and its strength.}

\be S={- {1\over 4}}\int d^{4}x\sqrt{-g}F_{\mu\nu}F^{\mu\nu}. \ee
For simplicity, we ignore the back reaction. It means that $A_{\mu}$
is a small perturbation and the metric is still the same as
\eqn{5eq5}.
This vector field is expected to support a charge current operator
$J_\mu$ in the dual field. The equation of motion of $A_\mu$ is \be
\frac{1}{\sqrt{-g}}\partial_{\mu}(\sqrt{-g}F^{\mu\nu})=0 \ee If we
only consider the zero component of $A_\mu$, $A=\phi(r)dt$, then we
have \be \phi''+\frac{z-1}{r}\phi'=0\;. \ee Near the boundary, \be
\phi=\phi_{(0)}+\phi_{(1)}r^{2-z}\;, \ee where $\phi_{(0)}$ and
$\phi_{(1)}$ are chemical potential and charge density respectively
in the dual field theory if $z<2$. 
In the special case $z=2$, 
\be\label{A0}
A_0=\mu_0 -\rho \log r/r_*. 
\ee 
The coefficient $\rho$ of the log term is precisely the charge density, which in grand canonical ensemble,  is defined as the derivative of the 
boundary action term with respect to the  chemical potential $\mu_0$.
See also \cite{Hartnoll:2009ns}  for related discussion. 

\section{Superconductivity}
In this section, we will build an Abelian-Higgs
model~\cite{Gubser:2008px,Hartnoll:2008vx} in the Lifshitz black
hole background and study the superconductive phase. We introduce a
new $U(1)$ gauge field $ A_\mu$ which is different from that in the
action (6) and also introduce a complex scalar $\psi$. We assume
that the background response is negligible for simplicity.
\subsection{Superconductive phases}
Considering the Lagrangian density
\begin{equation} \label{2eq2}
\mathcal{L}=-\frac{1}{4}F^{\mu\nu}F_{\mu\nu}-|\nabla\psi-iA\psi|^2-V(|\psi|),
\end{equation}we have equations of motion for $A$ and $\psi$ are respectively
\begin{equation} \label{2eq3}
\frac{1}{\sqrt{-g}}\partial_{\mu}(\sqrt{-g}F^{\mu\nu})=iq[\psi^*(\partial^{\nu}-iqA^{\nu})\psi
-\psi(\partial^{\nu}+iqA^{\nu})\psi^*],
\end{equation}
\begin{equation} \label{2eq4}
\frac{1}{\sqrt{-g}}\partial_{\mu}(\sqrt{-g}(\partial^{\mu}\psi-iqA^{\mu}\psi))-iqA^{\mu}
(\partial_{\mu}\psi-iqA_{\mu}\psi)-\frac{\psi}{2|\psi|}V'(|\psi|)=0.
\end{equation}
We will work in the probe limit, in which $A_{\mu}$ and $\psi$ are
taken to be small so that their back reaction on the spacetime
metric can be ignored. The metric is still a 4D Lifshitz black hole
with z=2 in \eqn{5eq5}.

Taking the ansatz $A=\phi(r)dt$, $\psi=\psi(r)$, these equations of
motion  reduce to
\begin{equation} \label{2eq5}
\phi''+\frac{z-1}{r}\phi'-\frac{2\psi^2}{r^2f(r)}\phi=0,
\end{equation}
\begin{equation} \label{2eq6}
\psi''+\left[\frac{f'(r)}{f(r)}-\frac{z+1}{r}\right]\psi'+\frac{r^{2z-2}\phi^2}{f^2(r)}\psi-\frac{V'(\psi)}{2r^2f(r)}=0.
\end{equation}
where $\psi$ can be taken to be real which is allowed by the r-component
of \eqn{2eq3}.
\begin{figure}\centering
  \includegraphics*[width=0.6\columnwidth]{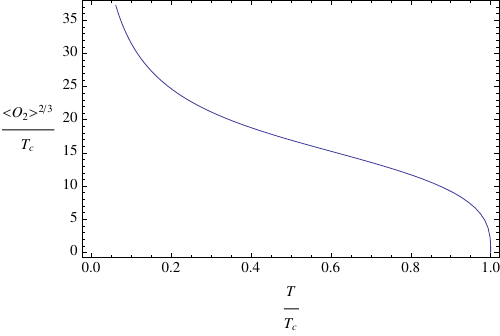}
  \caption{\small Condensation curve at $z=2, m^2=-3$. $<O_2>=\psi_{(1)}$.}\label{C}
\end{figure}
For simplicity we will specialize to a simple potential
$V(\psi)=m^2|\psi|^2$ with $m^2<0$ but above the
Breitenlohner-Freedman bound. Then near the boundary $r\rightarrow0$
the bulk fields behave as
\begin{equation} \label{2eq7}
\phi=\mu+\rho r^{2-z}\dots
\end{equation}
\begin{equation} \label{2eq8}
\psi=\psi_{(0)}r^{\nu_{-}}+\psi_{(1)}r^{\nu_{+}}+\dots
\end{equation}
with
$\nu_{\pm}=\frac{z+2}{2}\pm\sqrt{m^2+\left(\frac{z+2}{2}\right)^2}$.
At the horizon $\phi(r_H)=0$ and \eqn{2eq6} implies
\be
\psi'(r_H)=-{m^2\over 2r_H}\psi(r_H)\ .
\ee
For $z=2$, as we mentioned before, there is a Log sigularity for the second term in the right hand side of (\ref{2eq7}). We will study the regularized on-shell action in the following. Including $\phi(r)$ and $\psi(r)$, the bulk action can be rewritten as 
\be
S_{bulk}= V_3\int dr \sqrt{-g} \left[-{1\over 2}g^{rr}g^{tt}(\partial_r\phi)^2 - g^{rr}(\partial_r\psi)^2 - g^{tt}\phi^2\psi^2 - m^2\psi^2\right]
\ee
After doing the above integral by part and using the equations of motion, we have
\be\label{onshellaction}
S_{on-shell}= V_3\left[\sqrt{-g}\phi(-{1\over 2}g^{rr}g^{tt}\partial_r\phi)\biggr|_{r_B}^{r_H} +\sqrt{-g}\psi ( - g^{rr}\partial_r\psi)\biggr|_{r_B}^{r_H} +\int_{r_B}^{r_H}\sqrt{-g} g^{tt}\phi^2\psi^2\right]
\ee

Actually, the properties of boundary behaviors of  all three terms in (\ref{onshellaction}) heavily depend on the parameter $\nu_{\pm}$ related to $m$. In our discussion, the asymptotical behaviors of the second and the third term in (\ref{onshellaction}) are suppressed by $\psi(r)$. Using the asymptotical behaviors of $\phi(r)$ and $\psi(r)$, with eq. (\ref{2eq7}) replaced by 
\be
\phi(r) =  \mu_0-\rho \log {r\over r_*} +\cdots ,\ee 
 the
 action can be given by
\be
S = S_{on-shell}  = V_3\left[-{1\over 2}\rho(\mu_0 -\rho \log{\epsilon\over r_*}) +\cdots\right]\ ,
\ee
where $r_*$ was introduced in (\ref{A0}). 
Finally the regulated Euclidean total action of boundary field theory is given by
\be\label{onshellaction2}
S = {V_2\over T}\left(-{1\over 2}\rho\mu_{sc} + \int_{r_B}^{r_H}\sqrt{-g} g^{tt}\phi_c^2\psi_c^2\right)\ ,
\ee where $\mu_{sc}=\mu_0$.  We need to integrate the classical solution $\phi_c$ and $\psi_c$. The free energy is obtained by Legendre transformation: 
\be
F_{sc} =TS +\mu_{sc}\rho V_2 = V_2 ({1\over 2}\rho\mu_{sc} + \int_{r_B}^{r_H}\sqrt{-g} g^{tt}\phi_c^2\psi_c^2)\ .
\ee
For the normal state, the free energy is given by setting $\psi_c=0$ in (\ref{onshellaction2})
\be
F_{n} = V_2({1\over 2}\rho\mu_n) = V_2({1\over 2}\rho^2\log{r_H\over r_*})\ ,
\ee where we use the horizon regularity  condition $A_0(r_H)=0$ and $\mu_{n}=\rho\log{r_H\over r_*}$. 
\begin{figure}\centering
\includegraphics*[width=0.6\columnwidth]{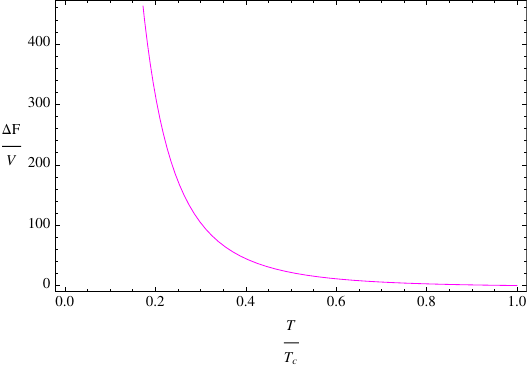}
 \caption{\small Difference of free energy curve with and without condensation,   $\Delta F=F_n-F_{sc}$ . }\label{fenergyd}
\end{figure}

 In the case with condensation, 
$\nu_{\pm}$  is simplified to $\nu_{-}=1$, $\nu_{+}=3$ with $m^2=-3$.  Note that  $\int_{\epsilon}^{r_H}\sqrt{-g} g^{tt}\phi_c^2\psi_c^2$ has a finite value.
 
The
condensate of the scalar operator $\mathcal{O}$ is encoded in the
dual field $\psi$ by
\begin{equation} \label{2eq9}
\langle\mathcal{O}\rangle=\psi_{(1)}
\end{equation}
with the boundary condition $\psi_{(0)}=0$. We can solve the equations \eqn{2eq5} and
\eqn{2eq6} numerically and finally get a condensation curve shown in
figure 1. Near the critical temperature,  this curve is similar to that in BCS theory and that in
z=1 holographic superconductor~\cite{Hartnoll:2008vx}.
$\langle\mathcal{O}\rangle$ goes to a finite value as the
temperature turns below a critical value. By dimensional analysis, $T_c\sim \mu$. 

We plot the difference of free energy  difference 
\be
\Delta F=F_n- F_{sc}   = V_2 \left({1\over 2}\rho(\mu_n-\mu_{sc}) + \int_{\epsilon}^{r_H}\sqrt{-g} g^{tt}\phi_c^2\psi_c^2\right)\ .
\ee
where $\mu_n, \mu_{sc}$ mean chemical potential at the normal and superconducting phase respectively.
Figure \ref{fenergyd} demonstrates that the  free energy for superconducting state is lower below critical temperature.

\subsection{Conductivity}
In order to compute the electric conductivity, we follow the
  procedure in \cite{Hartnoll:2009sz}. In this
paper we work in the probe limit so that the fluctuation of the metric or the massive gauge field  is
ignored.  For the conductivity only fluctuation $A_x(r)$ is relevant and let's  work in the zero spatial momentum  limit. 
Together with the background $\phi$ and the fluctuation $A_x$, 
\begin{equation} \label{3eq1}
A=\phi(r)dt+A_x(r)e^{-i\omega t}dx.
\end{equation}
From  the Maxwell equation,
\begin{eqnarray} \label{3eq3}
\frac{1}{\sqrt{-g}}\partial_{\mu}(\sqrt{-g}F^{\mu
x})=2\psi^2A^{x},\,\,\,\,\,\,
\end{eqnarray}
 we find
\begin{equation} \label{3eq7}
A_x''+\left[\frac{f'(r)}{f(r)}-\frac{z-1}{r}\right]A_x'+\left[\frac{\omega^2r^{2z-2}}{f^2(r)}-
\frac{2\psi^2}{r^2f(r)}\right]A_x=0.
\end{equation}

At the horizon we choose the infalling boundary condition,
\begin{equation} \label{3eq8}
A_x\propto f(r)^{- i\omega r_H^z/2}.
\end{equation}
Near the boundary,  the field behaves as
\begin{equation} \label{3eq9}
A_x=A_{x(0)}+A_{x(1)}r^z+\dots,
\end{equation}
where $A_{x(0)}$ gives the background electric field in the dual
field theory $E_x=i\omega A_{x(0)}$ and $A_{x(1)}$ is related to the
expectation of electric current $J_x$.

For the gauge field \eqn{3eq1}, the
Maxwell action reduces to
\begin{equation} \label{3eq10}
S=-\frac{2}{4}\int
d^4x\sqrt{-g}\left[g^{rr}(g^{xx}A_x'^2+g^{tt}\phi'^2)-g^{tt}g^{xx}\omega^2
A_x^2\right],
\end{equation}
then the expectation value of the electric current can be obtained
from this action,
\begin{equation} \label{3eq11}
\langle J^x\rangle=\frac{\delta S_{on-shell}}{\delta
A_{x(0)}}=-\lim_{r\rightarrow 0}\frac{\delta S}{\delta \partial_r A_{x}},
\end{equation}
with the notation $\partial_r A_{x }=A_{x}'(r)$. Finally, we
obtain
\begin{equation} \label{3eq12}
\langle J^x\rangle=\lim_{r\rightarrow
0}\sqrt{-g}g^{rr}(g^{xx}A_x')=zA_{x(1)},
\end{equation}
which gives the conductivity
\begin{equation} \label{3eq12}
\sigma(\omega)=\frac{\langle
J^x\rangle}{E_x}=-\frac{i}{\omega}\frac{zA_{x(1)}}{A_{x(0)}}.
\end{equation}
All left is to solve equation \eqn{3eq7} in order to obtain the
electric conductivity in \eqn{3eq12}.  In
particular, we plot the conductivity at $T<T_c$. Near $\omega=0$, we
observed a pole for the imaginary part. It means that DC conductivity becomes a delta function when
condensation happens. 

\begin{figure}\centering
  \includegraphics*[width=0.42\columnwidth]{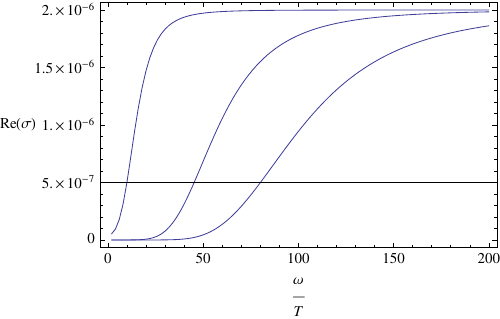}
\includegraphics*[width=0.4\columnwidth]{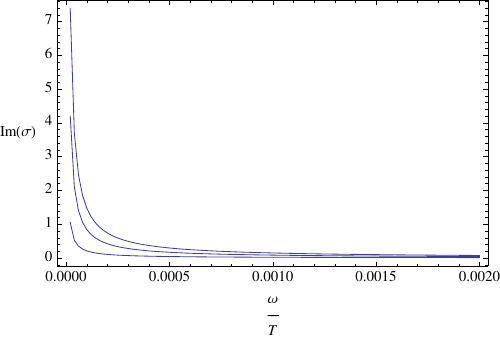}
  \caption{\small Conductivity at $T<T_c$.  The temperature $T=0.512T_c$ , $T=0.155T_c$ , and $T=0.090T_c$ from up to down in left figure and opposite in the right one.}\label{C3}
\end{figure}
Apparently, there is a gap in Fig.\ref{C3}. However, as pointed out in ~\cite{Horowitz:2009ij}, they may not be the genuine gap. One way to see this is to work out the the real part of the conductivity at low frequency at the zero temperature with full consideration of the backreaction to the gravity background. For charged Lifshitz solution (analogous to RN), we refer to~\cite{Brynjolfsson:2010rx}. Furthermore, the zero temperature limit from finite temperature is subtle, since a part of solutions at finite temperature depend on T only by the combination $\omega/T$. Therefore we can not take limit T goes to zero from the finite temperature.
We postpone the examination of the zero temperature limit separately to future investigation.

\section*{Acknowledgements:}
We acknowledge Wei-Shui Xu, Li-Ming Cao, Yan Liu, Zhao-long Wang and
Jian-Feng Wu, Shuo Yang for useful discussions. Shan-Shan Xu would
like to express thanks to J.X.Lu for his encouragement. Y.Zhou shall
thank his advisor Miao Li for encouraging this work. The work of SJS
was  supported by KOSEF Grant R01-2007-000-10214-0 and also by  NRF
grant  through CQUeST   with grant number 2005-0049409.


\begin{thebibliography}{99}

\bibitem{Maldacena:1997re}
J.~M.~Maldacena, ``The large N limit of superconformal field theories and supergravity,'' Adv.\ Theor.\ Math.\ Phys.\
{\bf 2} (1998) 231 [Int.\ J.\ Theor.\ Phys.\  {\bf 38} (1999) 1113] [arXiv:hep-th/9711200].

\bibitem{Polyakov:2001af}
  A.~M.~Polyakov,
  Int.\ J.\ Mod.\ Phys.\  A {\bf 17S1}, 119 (2002)
  [arXiv:hep-th/0110196].

\bibitem{Wilson:1974}
K. G. Wilson Phys. Rev. D10, 2445 (1974)

\bibitem{son}
G. Policastro, D. T. Son and A. O. Starinets,
 {\bf Phys. Rev. Lett.} {87}(2001) {081601}, hep-th/0104066.

\bibitem{SZ}
S.~J.~Sin and I.~Zahed, 
Phys. Lett. {B608}{(2005)}{265}, hep-th/0407215;

E. Shuryak, S-J. Sin and I. Zahed,
``A Gravity Dual of RHIC Collisions,''
J.Korean Phys. Soc. {\bf 50}, 384 (2007), hep-th/0511199.

\bibitem{Mateos:2007ay}
  For recent review,  see \\
  D.~Mateos,
  ``String Theory and Quantum Chromodynamics,''
  Class.\ Quant.\ Grav.\  {\bf 24}, S713 (2007)
  [arXiv:0709.1523 [hep-th]].

\bibitem{Hartnoll:2007ai}
  S.~A.~Hartnoll and P.~Kovtun,
  ``Hall conductivity from dyonic black holes,''
  Phys.\ Rev.\  D {\bf 76}, 066001 (2007)
  [arXiv:0704.1160 [hep-th]].

\bibitem{Hartnoll:2007ih}
  S.~A.~Hartnoll, P.~K.~Kovtun, M.~Muller and S.~Sachdev,
  ``Theory of the Nernst effect near quantum phase transitions in condensed
  matter, and in dyonic black holes,''
  Phys.\ Rev.\  B {\bf 76}, 144502 (2007)
  [arXiv:0706.3215 [cond-mat.str-el]].

\bibitem{Hartnoll:2007ip}
  S.~A.~Hartnoll and C.~P.~Herzog,
  ``Ohm's Law at strong coupling: S duality and the cyclotron resonance,''
  Phys.\ Rev.\  D {\bf 76}, 106012 (2007)
  [arXiv:0706.3228 [hep-th]].

\bibitem{Hartnoll:2008hs}
  S.~A.~Hartnoll and C.~P.~Herzog,
  ``Impure AdS/CFT,''
  arXiv:0801.1693 [hep-th].

\bibitem{Hartnoll:2008vx}
  S.~A.~Hartnoll, C.~P.~Herzog and G.~T.~Horowitz,
  ``Building a Holographic Superconductor,''
  Phys.\ Rev.\ Lett.\  {\bf 101}, 031601 (2008)
  [arXiv:0803.3295 [hep-th]].

\bibitem{Hartnoll:2008kx}
  S.~A.~Hartnoll, C.~P.~Herzog and G.~T.~Horowitz,
  ``Holographic Superconductors,''
  JHEP {\bf 0812}, 015 (2008)
  [arXiv:0810.1563 [hep-th]].

\bibitem{Ammon:2009fe}
  M.~Ammon, J.~Erdmenger, M.~Kaminski and P.~Kerner,
  ``Flavor Superconductivity from Gauge/Gravity Duality,''
  arXiv:0903.1864 [hep-th].

\bibitem{Horowitz:2008bn}
  G.~T.~Horowitz and M.~M.~Roberts,
  ``Holographic Superconductors with Various Condensates,''
  Phys.\ Rev.\  D {\bf 78}, 126008 (2008)
  [arXiv:0810.1077 [hep-th]].

\bibitem{Fujita:2009kw}
  M.~Fujita, W.~Li, S.~Ryu and T.~Takayanagi,
  ``Fractional Quantum Hall Effect via Holography: Chern-Simons, Edge States,
  arXiv:0901.0924 [hep-th].

\bibitem{Herzog:2008wg}
  D.~T.~Son,
  ``Toward an AdS/cold atoms correspondence: a geometric realization of the
  Schroedinger symmetry,''
  Phys.\ Rev.\  D {\bf 78}, 046003 (2008)
  [arXiv:0804.3972 [hep-th]].
  A.~Adams, K.~Balasubramanian and J.~McGreevy,
  ``Hot Spacetimes for Cold Atoms,''
  JHEP {\bf 0811} (2008) 059
  [arXiv:0807.1111 [hep-th]].
  C.~P.~Herzog, M.~Rangamani and S.~F.~Ross,
  ``Heating up Galilean holography,''
  JHEP {\bf 0811}, 080 (2008)
  [arXiv:0807.1099 [hep-th]].

\bibitem{Kachru:2008yh}
  S.~Kachru, X.~Liu and M.~Mulligan,
  Phys.\ Rev.\  D {\bf 78}, 106005 (2008)
  [arXiv:0808.1725 [hep-th]].

\bibitem{Mann:2009yx}
  R.~B.~Mann,
  ``Lifshitz Topological Black Holes,''
  JHEP {\bf 0906}, 075 (2009)
  [arXiv:0905.1136 [hep-th]].

\bibitem{Bertoldi:2009vn}
  G.~Bertoldi, B.~A.~Burrington and A.~Peet,
  ``Black Holes in asymptotically Lifshitz spacetimes with arbitrary critical exponent,''
  arXiv:0905.3183 [hep-th].

\bibitem{Bertoldi:2009dt}
  G.~Bertoldi, B.~A.~Burrington and A.~W.~Peet,
  ``Thermodynamics of black branes in asymptotically Lifshitz spacetimes,''
  arXiv:0907.4755 [hep-th].

\bibitem{Danielsson:2009gi}
  U.~H.~Danielsson and L.~Thorlacius,
  ``Black holes in asymptotically Lifshitz spacetime,''
  JHEP {\bf 0903}, 070 (2009)
  [arXiv:0812.5088 [hep-th]].

\bibitem{Brynjolfsson:2009ct}
  E.~J.~Brynjolfsson, U.~H.~Danielsson, L.~Thorlacius and T.~Zingg,
  ``Holographic Superconductors with Lifshitz Scaling,''
  arXiv:0908.2611 [hep-th].

\bibitem{Taylor:2008tg}
  M.~Taylor,
  ``Non-relativistic holography,''
  arXiv:0812.0530 [hep-th].

\bibitem{Balasubramanian:2009rx}
  K.~Balasubramanian and J.~McGreevy,
  ``An analytic Lifshitz black hole,''
  arXiv:0909.0263 [hep-th].

\bibitem{AyonBeato:2009nh}
  E.~Ayon-Beato, A.~Garbarz, G.~Giribet and M.~Hassaine,
  ``Lifshitz Black Hole in Three Dimensions,''
  arXiv:0909.1347 [hep-th].

\bibitem{Cai:2009ac}
  R.~G.~Cai, Y.~Liu and Y.~W.~Sun,
  ``A Lifshitz Black Hole in Four Dimensional $R^2$ Gravity,''
  arXiv:0909.2807 [hep-th].

\bibitem{Li:2009pf}
  W.~Li, T.~Nishioka and T.~Takayanagi,
  ``Some No-go Theorems for String Duals of Non-relativistic Lifshitz-like Theories,''
  arXiv:0908.0363 [hep-th].

\bibitem{Hartnoll:2009sz}
  S.~A.~Hartnoll,
  ``Lectures on holographic methods for condensed matter physics,''
  arXiv:0903.3246 [hep-th].

\bibitem{Son:2002sd}
  D.~T.~Son and A.~O.~Starinets,
  ``Minkowski-space correlators in AdS/CFT correspondence: Recipe and
  applications,''
  JHEP {\bf 0209}, 042 (2002)
  [arXiv:hep-th/0205051].

\bibitem{Gubser:2008px}
  S.~S.~Gubser,
  ``Breaking an Abelian gauge symmetry near a black hole horizon,''
  arXiv:0801.2977 [hep-th].

\bibitem{Aizawa:2009yc}
  N.~Aizawa and V.~K.~Dobrev,
  arXiv:0906.0257 [hep-th].
\bibitem{Konoplya:2009hv}
  R.~A.~Konoplya and A.~Zhidenko,
  arXiv:0909.2138 [hep-th].

\bibitem{Koroteev:2007yp}
  P.~Koroteev and M.~Libanov,
  JHEP {\bf 0802}, 104 (2008)
  [arXiv:0712.1136 [hep-th]].

\bibitem{Pu:2009wn}
  S.~Pu, S.~J.~Sin and Y.~Zhou,
  arXiv:0903.4185 [hep-th].

\bibitem{Hartnoll:2009ns}
  S.~A.~Hartnoll, J.~Polchinski, E.~Silverstein and D.~Tong,
  JHEP {\bf 1004}, 120 (2010)
  [arXiv:0912.1061 [hep-th]].

\bibitem{Horowitz:2009ij}
  G.~T.~Horowitz and M.~M.~Roberts,
  JHEP {\bf 0911}, 015 (2009)
  [arXiv:0908.3677 [hep-th]].
  
\bibitem{Brynjolfsson:2010rx}
  E.~J.~Brynjolfsson, U.~H.~Danielsson, L.~Thorlacius and T.~Zingg,
  JHEP {\bf 1008}, 027 (2010)
  [arXiv:1003.5361 [hep-th]].
  E.~J.~Brynjolfsson, U.~H.~Danielsson, L.~Thorlacius and T.~Zingg,
  arXiv:1004.5566 [hep-th].

\end{thebibliography}
\end{document}